\documentclass[10pt]{iopart}

\usepackage{graphicx}%
\usepackage{multirow}%
\usepackage{amsmath}%
\usepackage{xcolor}%
\usepackage{booktabs}%

\usepackage[style=nature]{biblatex}%
\begin{document}

\title[The Radiation Pressure of Light]
{The Radiation Pressure of Light: historical perspectives and the role of structured light}

\author{Alexander C. Trowbridge}
\address{Centre of Light for Life, School of Biological Sciences, The University of Adelaide, Adelaide, 5005, SA, Australia}
\address{Institute for Photonics and Advanced Sensing, The University of Adelaide, Adelaide, 5005, SA, Australia}

\author{Ewan M. Wright}
\address{Wyant College of Optical Sciences, The University of Arizona, 1630 East University Boulevard, Tucson, AZ, 85721, USA}
\address{SUPA, School of Physics \& Astronomy, University of St Andrews, North Haugh, St Andrews, KY16 9SS, UK}

\author{Kishan Dholakia}
\address{Centre of Light for Life, School of Biological Sciences, The University of Adelaide, Adelaide, 5005, SA, Australia}
\address{Institute for Photonics and Advanced Sensing, The University of Adelaide, Adelaide, 5005, SA, Australia}
\address{SUPA, School of Physics \& Astronomy, University of St Andrews, North Haugh, St Andrews, KY16 9SS, UK}
\address{Wyant College of Optical Sciences, The University of Arizona, 1630 East University Boulevard, Tucson, AZ, 85721, USA}



\ead{kishan.dholakia@adelaide.edu.au}

\vspace{10pt}
\begin{indented}
\item[]January 2025
\end{indented}

\begin{abstract}
Light, or electromagnetic radiation, is well known to possess momentum, and the exchange of this momentum with a reflecting surface leads to radiation pressure. More often than not, it is the radiation pressure generated by a plane wave incident on a flat mirror that is considered. The last few decades have seen the emergence of structured light beams that may possess a complex phase and amplitude structure in both their transverse and longitudinal directions. This paper provides a historical overview of radiation pressure, tracing its discovery and experimental validation, and examines how transitioning to structured light from a plane wave can influence it. In particular, we elucidate the difference in radiation pressure force for structured light fields and how this differs from that of a plane wave at an identical frequency. In particular, the well-known Gouy phase is shown to contribute to a reduction in the radiation pressure force exerted on a flat mirror in comparison to a plane wave for both HG and LG modes. As an illustrative example, we compute that the radiation pressure force for LG modes differs from that of a plane wave by approximately $20$ fN/W for each unit of orbital angular momentum. A detailed experimental proposal to quantify this variance in radiation pressure is described, and we demonstrate that this measurement is within the realm of current metrological techniques.
\end{abstract}

\vspace{2pc}
\noindent{\it Keywords}: Radiation Pressure, Structured Light, Gouy Phase, Orbital Angular Momentum


%
\ioptwocol

\section{Introduction}
When electromagnetic radiation is incident on a surface, it exerts a pressure termed the radiation pressure. Radiation pressure has fascinated scientists since Maxwell's calculations in 1873, which form the basis for optical manipulation of mesoscopic particles and atoms \cite{simpson_mechanical_1997, chen_dynamics_2013, padgett_orbital_2017, shaw_comparison_2019, lembessis_mechanical_2016}. It can be determined by considering either the momentum of a classical electromagnetic field or in terms of photon momenta. The majority of studies in this area have focused on infinite-plane waves, and this has been of central importance in the fields of atom and particle micromanipulation.

In this paper we consider the radiation pressure of finite, structured light beams and how we may measure this pressure for such beams when compared to plane waves of the same frequency. Structured light beams include transverse laser modes such as Hermite-Gaussian and Laguerre-Gaussian beams, as well as other mode families \cite{davis_theory_1979}. We find that such structured light fields exhibit a slight reduction in radiation pressure compared to their plane wave counterparts, on the order of femtoNewtons. The origin of this difference lies in the Gouy phase of the light field, which is the accumulation of an additional phase shift compared to a plane wave propagating over the same distance \cite{subbarao_topological_1995,feng_physical_2001}. In this study, we describe how the optical radiation pressure force is influenced by the mode structure via the Gouy phase shift.

As an illustrative example, we consider the case of Laguerre-Gaussian (LG) modes and describe explicit experimental manifestations to measure this radiation pressure difference relative to a plane wave, which is of magnitude 19 fN/W per quanta of orbital angular momentum (azimuthal index). A similar magnitude of pressure exists for the indices from Hermite-Gaussian (HG) modes. It is thought that, based on the predictions of this model, a direct measurement of the pressure reduction from beams with angular momentum may be possible. Force measurements of this scale have been demonstrated in numerous studies \cite{shaw_comparison_2019, li_traceable_2019, qiu_directly_2021, leach_direct_2006, lembessis_mechanical_2016, rohrbach_switching_2005}. An experiment is proposed to realize this measurement.

\section{History of Radiation Pressure}
In the 16th century, astronomy was deeply intertwined with the sociopolitical and religious structures of the time. Scientific inquiry operated within this framework, often under noble patronage, and was aligned with theological worldviews. Tycho Brahe, the renowned Danish astronomer, exemplified this intersection. His observation of the Great Comet of 1577, meticulously recorded over several months, challenged prevailing Aristotelian cosmology, which held that celestial objects, particularly comets, were atmospheric phenomena bound within the Earth’s sphere \cite{christianson_tycho_1979}. By tracking the comet’s path, Brahe demonstrated that it was far beyond the Earth’s atmosphere, travelling through the “celestial” regions. Particularly noteworthy in the context of radiation pressure, Brahe’s observations included his record that the comet’s tail always pointed away from the Sun. This consistent orientation suggested the action of an unknown force affecting celestial bodies, an idea that transcended contemporary understanding but began to reshape the framework of celestial mechanics \cite{christianson_tycho_1979}.

Following Brahe, Johannes Kepler expanded upon these ideas with further ambition. In the early 17th century, Kepler theorised that a force originating from the Sun was responsible for influencing the movement of planetary bodies and celestial objects, including comets. He proposed that this “solar force” was responsible for directing comet tails away from the Sun, essentially aligning with what would later be understood as radiation pressure. Kepler’s concept of radiation pressure was speculative and rooted in an intuitive grasp of the forces emanating from the Sun, yet it bore a substantial resemblance to future scientific descriptions. In his work *Astronomia Nova* (1609), Kepler posited that sunlight might exert a tangible pressure, which could physically influence the trajectories of comets and even planets. While Kepler’s ideas were intertwined with his metaphysical beliefs about celestial harmonies, his intuition about a solar force influencing matter in space marked a profound departure from purely geocentric explanations, guiding astronomy toward a heliocentric understanding of gravitational and non-gravitational forces.

Kepler’s hypotheses on solar radiation, though lacking empirical confirmation, resonated through the scientific community and gradually drew interest toward the question of light as a physical entity with potential momentum. These early ideas were largely speculative, but they revealed a critical shift in thinking about light and its interactions with matter. The notion that sunlight could influence objects in space awaited formalisation until James Clerk Maxwell’s development of electromagnetic theory in the 19th century. Although multiple experimental attempts were made to test the theory in the 18th century, they met with little success. Kepler's hypotheses set the stage for later scientific enquiry, culminating in Maxwell's unification of electromagnetic theory, which provided the first quantitative framework for radiation pressure. Maxwell’s equations offered a groundbreaking revelation: light, as an electromagnetic wave, carries both energy and momentum, implying that it could exert a measurable pressure upon objects. Known as the Maxwell-Bartoli theory, this prediction transformed light from a purely passive medium into an active participant in physical interactions. For a given incident intensity $I$, and surface reflectivity $\rho$, the Maxwell-Bartoli equation for the photon pressure is

\begin{equation} \label{eq:maxwell_bartoli}
    \text{Photon Pressure} = \frac{I(1+\rho)}{c} ,
\end{equation}

and for a perfectly reflecting flat mirror this yields the radiation pressure force $F=2P/c$.

In 1865, James Clerk Maxwell’s groundbreaking theory of electromagnetism provided not only a unification of electric and magnetic fields but also a series of profound predictions, one of which was the phenomenon of radiation pressure. Maxwell’s equations implied that light, as an electromagnetic wave, could exert force on matter. This insight was theoretically extended by Bartoli in 1876, who, using principles of thermodynamics, inferred that radiation pressure could perform work, establishing light as a quantifiable force-carrier. Maxwell’s formalism thus suggested that a beam of light impacting a surface could impart a minute, yet measurable, pressure – a prediction that sparked an intense experimental drive to validate this effect.

Measuring such a subtle force presented subtle technical challenges, as the predicted pressures were extraordinarily small, necessitating instruments with high sensitivity and stability. Early attempts to confirm Maxwell’s prediction were plagued by experimental limitations, especially the difficulty of isolating true radiation pressure from confounding thermal and convective forces, known as radiometric effects. This issue is clearly illustrated by the Crookes radiometer developed in 1873, the same year Maxwell published ``A Treatise on Electricity and Magnetism" and three years prior to Bartoli's thermodynamic work. The radiometer’s motion, originally thought to result from radiation pressure, was in fact driven by temperature differentials. By the turn of the 20th century, the scientific community was poised to resolve these challenges, with multiple experimental groups vying to be the first to achieve a definitive measurement.

\begin{figure}[htb!]
    \centering
    \includegraphics[width=\linewidth]{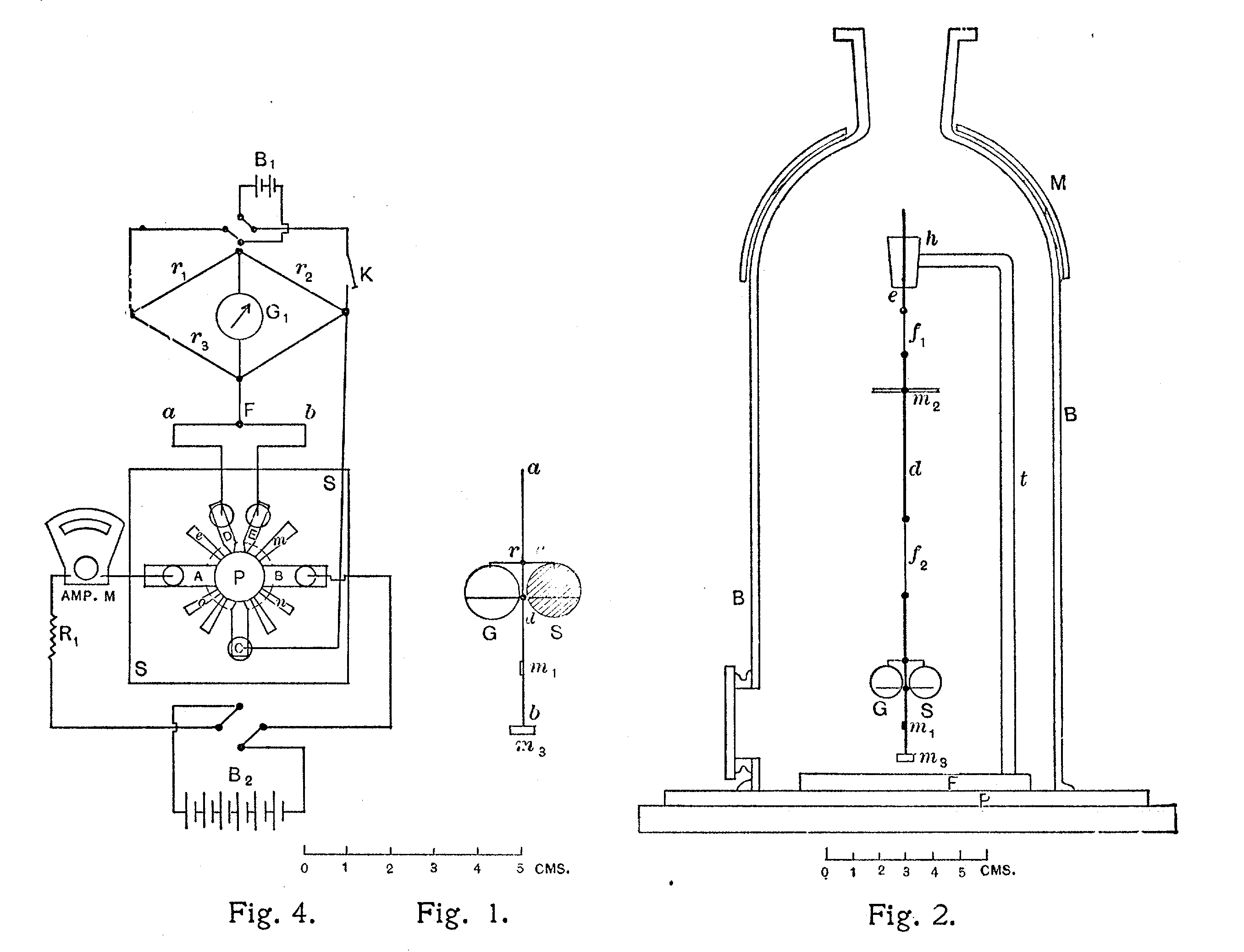}
\caption{The experimental diagram published by Nichols \& Hull in their landmark paper \textcite{nichols_preliminary_1901}. Like many others attempting to measure the Maxwell-Bartoli effect at the time, their radiometry apparatus was based around a torsion balance to achieve the fine sensitivity required for the measurement. Achieving suitable isolation from erroneous thermodynamic forces plagued the numerous measurement attempts. (reprinted with permission)}
    \label{fig:nichols_hull}
\end{figure}

The Russian physicist Pyotr Lebedev was among the earliest to achieve notable success. In his 1900 experiments, Lebedev demonstrated that light indeed exerted pressure on small, suspended plates, providing the first experimental confirmation of Maxwell’s radiation pressure \cite{lebedew_untersuchungen_1901}. In 1901, American physicists Ernest Fox Nichols and Gordon Ferrie Hull conducted their own series of experiments at Dartmouth College \cite{nichols_preliminary_1901}. Like Lebedev, Nichols and Hull aimed not only to confirm the existence of radiation pressure but also to rigorously test Maxwell’s quantitative predictions. Their approach, independent of Lebedev’s work, involved using a similar torsion balance to measure the minute forces exerted by light on reflective and absorptive surfaces. Depicted in Figure \ref{fig:nichols_hull}, Nichols and Hull's torsion balance experiment marked a milestone in measuring the Maxwell-Bartoli effect. Unaware of Lebedev's recently published work, Nichols and Hull also claimed that they were the first to experimentally quantify and prove the existence of the Maxwell-Bartoli effect.

In 1903, Nichols and Hull published a work which was, in part, both a response to and a critique of Lebedev’s work. In the paper, they openly acknowledged Lebedev’s pioneering role while positioning their own work as a more refined and quantitative investigation. They noted, for instance, that while Lebedev’s results aligned broadly with Maxwell’s predictions, the inherent “accidental errors” in his measurements limited their precision. Nichols and Hull’s apparatus, although initially flawed due to “false resistance” in their bolometer, eventually allowed them to make a more precise measurement that brought their findings closer to theoretical expectations \cite{nichols_pressure_1903}. In a carefully worded statement, they claimed that Lebedev’s experiment was sufficient to establish the existence of radiation pressure but that it fell short of providing a “satisfactory quantitative confirmation” of the Maxwell-Bartoli theory.

The nuanced stance of Nichols and Hull underscores the competitive spirit and high scientific standards of the era. Their work was not without controversy, especially among Lebedev’s supporters, as it questioned the precision of his results while acknowledging his priority in the discovery. Despite this, Nichols and Hull were careful to maintain a professional tone, noting that their initial publication had similarly lacked quantitative rigor due to early equipment issues. This statement reflects a delicate balance of scientific critique and respect, typical of the early 20th-century experimental physics community.

The experiments of Lebedev, Nichols, and Hull represented crucial steps in the historical trajectory of radiation pressure research, underscoring its evolution from theoretical prediction to experimental validation. Their contributions laid the groundwork for later advances in precision measurement and optical manipulation, leading to applications that Maxwell and Bartoli could scarcely have imagined. This early work not only confirmed radiation pressure as a real and measurable force but also set a standard for future investigations into the interactions between light and matter, a standard that experimental physicists continue to refine today.

Around the same time, Albert Einstein published his 1905 paper on the photoelectric effect, proposing the quantum nature of light as discrete packets, or photons. This work not only provided the foundation for quantum mechanics but also influenced the understanding of radiation pressure, as it suggested that photons themselves carry momentum \cite{einstein_uber_1905}. According to Einstein’s interpretation, the energy of a photon is directly proportional to its frequency, while its momentum is inversely proportional to its wavelength. This insight revolutionized how scientists viewed the interaction between light and matter, unifying Maxwell’s predictions with the emerging quantum perspective and providing a clearer pathway to interpreting radiation pressure within a quantum framework. Einstein’s contributions transformed theoretical physics, ultimately allowing subsequent experiments to validate the mechanical force of individual photons, a concept revisited and expanded by later studies in photon momentum and light’s angular properties. This quantum perspective naturally yields the uncertainty arguments important for describing the radiation pressure of structured light.

In 1909, John Henry Poynting further extended the theory of light-matter interactions by exploring the angular momentum carried by light. Poynting postulated that light with circular or helical polarization could transfer angular momentum to matter, exerting a torque upon absorption or reflection from a surface \cite{poynting_wave_1909}. However, the prevailing sentiment at the time – expressed explicitly by Poynting – was that the effect would be far too subtle to measure experimentally. This scepticism did not deter physicist Richard Beth, who in 1936 successfully demonstrated the transfer of angular momentum from light to a material system. \textcite{beth_mechanical_1936} used a quartz waveplate to measure the minuscule rotation induced by circularly polarized light, providing the first experimental confirmation of light’s angular momentum. His work not only validated Poynting’s predictions but also underscored the growing importance of light’s structural properties, foreshadowing modern applications where structured light, such as vortex beams, plays a crucial role. This development establishes a direct historical connection to the mechanical effects of structured light on matter.

As understanding of radiation pressure evolved, so too did its practical applications. One of the most ambitious applications was the concept of solar sails, which utilise radiation pressure from sunlight for propulsion. First proposed in the 20th century and later demonstrated in missions like JAXA’s IKAROS and the Planetary Society’s LightSail, solar sails highlight how accumulated radiation pressure over large surfaces can provide a sustainable propulsion method for deep-space exploration. Solar sails illustrate the potential of radiation pressure on a macroscopic scale, expanding the domain of optical forces to applications far beyond laboratory settings.

On a microscopic scale, Arthur Ashkin’s pioneering work in the 1970s and 1980s harnessed radiation pressure in a different context—optical trapping and manipulation. Ashkin’s optical tweezers, which use focused laser beams to trap and manipulate small particles, marked a revolutionary development in biological and physical research, enabling precise manipulation of cells, molecules, and even individual atoms. This work demonstrated the versatility and power of radiation pressure on microscopic particles, a discovery that earned him the Nobel Prize in Physics in 2018. His innovation underscores the profound impact of radiation pressure across different scales, from biological applications to fundamental studies in optics and mechanics. This progression highlights the potential for modern applications of radiation pressure, including structured light experiments examined later in this paper, where the angular and spatial characteristics of light are key to understanding and manipulating optical forces.

By bridging the historical trajectory of radiation pressure research with the nuanced behaviour of structured light, this work underscores the enduring relevance of foundational studies in addressing modern challenges in photonics.

\section{Structured Light Formalism}
Structured light refers to optical fields engineered to possess spatially varying properties—such as intensity, phase, polarization, or angular momentum—that differ from conventional Gaussian or plane waves. By carefully designing these spatial structures, one can tailor the light to interact with matter in new and nuanced ways, creating applications across optical trapping, imaging, communications, and quantum information. This concept has led to the development of beams with unique topological features, such as vortex beams carrying orbital angular momentum and vector beams possessing spatially modulated polarization. Structured light not only expands the toolbox of photonics but also introduces new ways to control radiation pressure, enabling forces that depend not only on the beam’s power but also on its specific structure. This section develops the formalism describing how the structure of light influences the corresponding radiation pressure, extending classical models to capture the unique interactions between these specialised beams and reflective or refractive surfaces.

\subsection{Radiation Pressure: the effect of structured light}
We represent the (normalized) complex electric field envelope for a monochromatic field propagating dominantly along the z-axis in bra-ket notation as $|E\rangle$, with $E(x,y,z) = \langle x,y,z | E\rangle$. Using this notation, the axial component of the field’s wavevector can be expressed as

\begin{equation}\label{kz}
\langle k_z \rangle = \langle E | \hat k_z  |E \rangle = \langle E \left |-i\partial_z \right | E \rangle ,  \quad \langle E|E\rangle = 1.
\end{equation}

For a perfectly reflecting and flat mirror, the radiation pressure force along the z-axis will be proportional to twice $\langle k_z \rangle$ for the incident field, being $2P/c$ in the limit of a plane wave.

Consider the general form of the complex electric field propagating dominantly along the z-axis

\begin{equation} \label{E}
    E(x,y,z) = A(x,y,z)e^{i\theta(x,y,z)},
\end{equation}

with complex amplitude $A(x,y,z)$, and where the phase $\theta(x,y,z)$ is given by

\begin{equation}\label{theta}
\theta(x,y,z) = kz + \frac{k(x^2+y^2)}{2R(z)} - (N+1)\arctan\left ( \frac{z}{z_0} \right ).
\end{equation}

The first term is due to the dominant plane-wave contribution to the wavevector, the second term accounts for the phase curvature with $R(z)$ being the radius of curvature, and the last term is the mode-dependent Guoy phase-shift $\theta_G(z) = - (N+1)\arctan( {z/z_0})$: here, $N=n+m$ for the case of HG modes, with $n$ and $m$ being the transverse mode indices, and $N=2p+|\ell|$ for LG modes, where $p$ is the radial mode index and $\ell$ the azimuthal mode index. It is known that the Guoy phase is a Berry phase, and thus is geometric in nature \cite{subbarao_topological_1995}.

Substituting the field in Eq. (\ref{E}) into Eq. \eqref{kz} we obtain

\begin{eqnarray}
\langle k_z \rangle &=& \int dx \int dy~A^* \left (-i\partial_zA + A \partial_z\theta \right ) \nonumber \\
&=& -i\int dx \int_{ }^{ } dy~A^* \partial_zA + \langle \partial_z\theta \rangle .
\end{eqnarray}

For the modal solutions of interest the first term is zero via the following argument - for HG modes we may choose $A(x,y,z)$ as real, so the first integral may be expressed as

\begin{equation}
\int dx \int dy~A^* \partial_zA  = \frac{d}{dz}\int dx \int dy~|A(x,y,z)|^2 = 0 ,
\end{equation}

which is zero due to power conservation. Similarly, for the LG modes, $A(x,y,z)$ has the azimuthal variation $e^{i\ell\phi}$, which does not vary with $z$, so the same argument applies. For the modal solutions of interest and the phase in Eq. (\ref{theta}), we then obtain

\begin{equation}\label{kz_final}
\langle k_z \rangle = \langle \partial_z\theta \rangle = k + \partial_z\theta_G +  \frac{1}{2}k \langle x^2+y^2\rangle~\partial_z R^{-1} .
\end{equation}

Working through the algebra, one can finds the axial component of the wavevector can be expressed as

\begin{equation} \label{kz_before}
    \langle k_z \rangle = k - (N+1)\frac{z_0}{z^2+z_0^2} + z_0\frac{\langle x^2+y^2 \rangle}{w_0^2}\frac{z^2-z_0^2}{(z^2+z_0^2)^2} .
\end{equation}

Feng and Winful \cite{feng_physical_2001} present an elegant solution for evaluating $\langle x^2+y^2 \rangle$ for HG beams. The approach hinges on the fact that photons have a well-defined total momentum, but not necessarily a well-defined direction, which can be used to provide an upper bound on the second moment of the axial direction of the wave vector. The same expression for spot size was computed by Carter \cite{carter_spot_1980}, and the corresponding result for LG beams has also been demonstrated \cite{phillips_spot_1983}.

The variance of position in each direction has the form $\Delta x\Delta k_x = \frac{1}{2} + m$. It can also be shown that $\langle k_x^2 \rangle = 2/w^2 (1/2+m)$. Combining these expressions yields the spot size in the $x$ direction, $\Delta x^2 = w^2(1/2+m)$. Similarly for $y$, $\Delta y^2 = w^2(1/2+n)$. Noting that the radial coordinate cannot become negative, and thus the variance picks up a factor of 1/2, the spot size becomes,

\begin{equation} \label{spot_size}
    \langle x^2+y^2 \rangle = \frac{1}{2}(1+m+n)w^2 = \frac{1}{2}(N+1)w^2.
\end{equation}

As the HG modes form a complete orthogonal set, this expression in terms of the total mode number holds for any beam. It will also hold for non-eigenmode light structures, except the total mode number then becomes the expected value of that quantity. This is evident as non-eigenmode beams must be expressible as a linear combination of eigenmodes, such as the Hermite-Gaussian basis. Each term can be treated independently as above, resulting in a weighted sum, which is, in effect, the expected value of the quantity.

\begin{equation}
    \langle k_z \rangle = k - \frac{N+1}{2z_0}
\end{equation}

It therefore follows that the radiation pressure normal to the beam depends on the structure of the light via the total mode number.

\begin{equation}\label{eq:rad_pressure}
    \frac{F_z}{2P/c} = 1 - \frac{N+1}{k^2w_0^2}
\end{equation}

The shift in the axial force is proportional to $N+1$, a dependence that aligns directly with the structure of the Gouy phase shift. While the Gouy phase is sometimes interpreted as a geometric phase, such an interpretation is not necessary to appreciate this connection. Readers familiar with beam optics will recognise that the mode-dependent Gouy phase intuitively mirrors the dependence of the axial force shift on the total mode number. This perspective highlights the elegance of structured light: its modal composition not only defines its spatial profile but also directly influences its optomechanical properties, such as radiation pressure.

Note that the ratio of the axial radiation force to the plane-wave value in Eq. (\ref{eq:rad_pressure}) is always less than unity, suggesting that the expected propagation speed of the wave in the axial direction is always marginally less than $c$. This is in agreement with previous studies regarding the propagation of structured light \cite{lembessis_mechanical_2016,leach_direct_2006,qiu_directly_2021,stoyanov_gouy_2023,wan_propagation_2023}. This force difference is extremely subtle. For a $632.8$ nm HeNe laser focused to a waist of $w_0 = 0.1$ mm, the difference is expected to be approximately one part per million ($10^{-6}P/c$) per unit mode number. Thus, the alignment of any experiment would have to be very precise to avoid masking this effect. At the waist, for small perturbations the wavefronts remain approximately parallel to a flat mirror surface. Thus, combined with conservation of momentum, no change in force consistent with deviation from perfect orthogonal alignment is expected. However, the force will change with small angular deviations from orthogonality. If the mirror is misaligned by too great an angle, such as one arcminute, the subtle radiation pressure effect will be masked. The angle at which this occurs can be approximated by expanding the cosine.

\begin{equation}
    \frac{F_z}{2P/c} = \cos\theta \simeq 1 - \frac{\theta^2}{2} \implies \theta \simeq \frac{\sqrt{2}}{kw_0}
\end{equation}

The maximum permissible angular deviation for the laser operating at 632.8 nm, for example, is approximately 0.5 arcminutes per unit $N$. An angular uncertainty of less than this value is required to ensure the effect is resolvable.

\subsection{Connection to the Beam Propagation Factor}
The radiation pressure correction depends largely on the mode number of the incident beam. This indicates a connection to the widely used beam propagation factor, $M^2$. For a perfect HG beam, this factor is $M^2 = m+n+1$, which, due to the orthogonality and superposition of modes, naturally integrates into the theory presented in this work \cite{saghafi_beam_1998}. The modified version of equation \eqref{eq:rad_pressure} is thus,

\begin{equation}
    \frac{\Delta F_z}{2P/c} = -\frac{M^2}{(kw_0)^2}.
\end{equation}

The denominator has a geometric connection; it is related to the diffraction angle, $\theta = w_0/z_0$. Using this relation, the radiation pressure expression can be written in terms of the $M^2$ factor and the diffraction angle.

\begin{equation}
    \frac{\Delta F_z}{2P/c} = -\frac{1}{4}M^2\theta^2
\end{equation}

There are a variety of techniques in which these factors can be measured \cite{siegman_how_1998, hyde_m2_2019, saghafi_beam_1998}, making this force potentially more accessible experimentally.

\subsection{Example: The Laguerre-Gaussian modes}
The radiation pressure of beams with orbital angular momentum (OAM) is of interest for optical trapping and manipulation experiments. A natural basis for OAM beams is the LG modes. Equation \eqref{eq:LG_mode} shows the complex amplitude of a general LG mode, normalized so that the integral over the infinite plane at a constant $z$ is unity, that is, $\langle E|E \rangle = 1$. Of course, since we know the formulation of the total mode number $N$, one could immediately write down the radiation pressure as per equation \eqref{eq:rad_pressure}, but this is not always the case.

\begin{multline}\label{eq:LG_mode}
    E(r,\phi,z) = C_{lp}\frac{1}{w(z)}\left(\frac{r\sqrt{2}}{w(z)}\right)^{|l|}\exp\left(-\frac{r^2}{w^2(z)}\right) \\ L_p^{|l|}\left(\frac{2r^2}{w^2(z)}\right)e^{il\phi}e^{-i\theta}
\end{multline}

Here, $C_{lp}$ is a normalisation coefficient of the form $\sqrt{p!/(p+|l|)!}$. $l$ and $p$ are the characteristic mode numbers, which give rise to the family of orthogonal complex amplitudes, $A_{lp}$. The force possesses azimuthal symmetry and is covered by the case of an axisymmetric beam.

\begin{equation}
    A_{lp}(r,\phi) = C_{lp}\left(\frac{\sqrt{2}r}{w}\right)^{|l|}L_p^{|l|}\left(\frac{2\rho^2}{w^2}\right)e^{-il\phi}
\end{equation}

Since the mode function is even about $l$, only the $l \ge 0$ case is considered during the calculations without loss of generality. Through some manipulation of the generalized Laguerre polynomials, it is possible to arrive at equation \eqref{eq:expanded_A}.

\begin{multline} \label{eq:expanded_A}
    \frac{2r^2}{w^2}A_{lp} = (2p+l+1)A_{lp} \\ - (p+1)\left(\frac{C_{lp}}{C_{lp-1}}A_{lp-1} + \frac{C_{lp}}{C_{lp+1}}A_{lp+1}\right)
\end{multline}

The familiar $2p+l+1$ term from the Gouy phase is readily observable. By orthogonality, all other modes vanish within the overlap integral when substituted into Equation \eqref{spot_size} to obtain the total mode number, $N$.

\begin{equation}
     \langle A_{lp}G | 2r^2/w^2 | A_{lp}G \rangle = 2p + l + 1 \implies N = 2p + l
\end{equation}

By symmetry, this implies that $N = 2p + |l|$ when generalised to all $l$. Substituting into equation \eqref{eq:rad_pressure} yields the closed-form expression for radiation pressure of an LG beam on an idealised planar surface. If the surface is an ideal mirror, the force gains a factor of two due to the reflection.

\begin{equation}\label{eq:LG_force}
     \frac{F_z}{(2P/c)} = 1 - \frac{2p + |l| + 1}{k^2w_0^2}
\end{equation}

\begin{figure}[htb!]
    \centering
    \includegraphics[width=\linewidth]{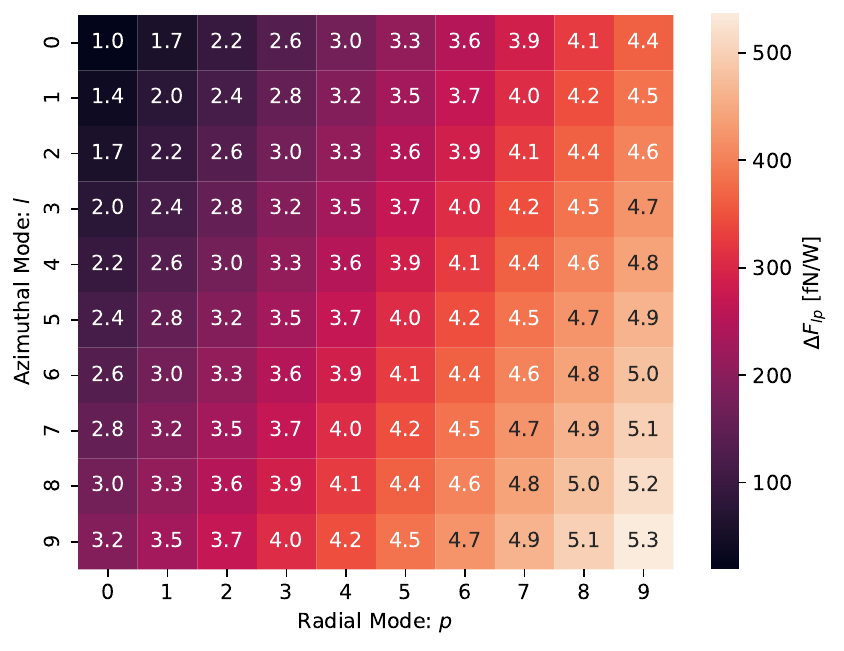}
\caption{Radiation pressure differences arising from LG modes with differing radial and azimuthal indices. The radiation pressure values follow from equation \eqref{eq:LG_force}, and the values within each square correspond to beam widths in units of $w_0$. The values are based on a 1064nm laser with a 0.1mm beam waist.}
    \label{fig:LG_modes}
\end{figure}

This expression is evaluated in Figure \ref{fig:LG_modes}, showing the magnitude and beam widths for each mode combination. The beam width and the strength of the effect scale proportionally. It can be seen that the force is on the order of $\sim 0.1$ pN/W for a $1064$ nm laser with a $0.1$ mm waist. For a $10$ W laser, this estimates the force to be $0.19$ pN per total mode number, $N$. The corresponding maximum permissible angular misalignment is $0.86$ arcminutes.

\section{Experimental Realisation}
\begin{figure*}[htb!]
    \centering
    \includegraphics[width=\textwidth]{"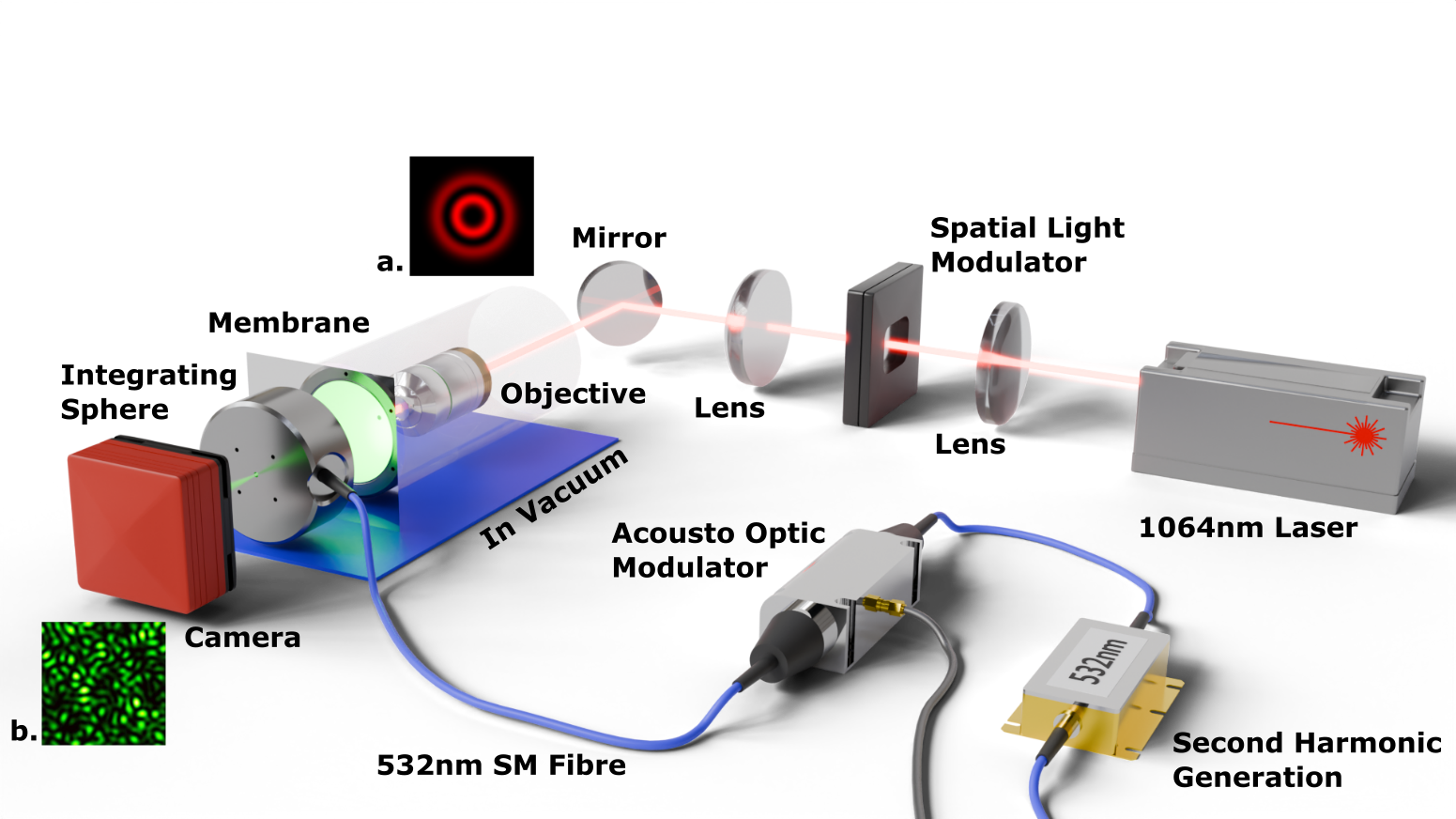"}
\caption{The optical layout of the proposed radiation pressure experiment. Here, a laser is shaped into an LG mode and focused onto a membrane via a spatial light modulator (SLM) and objective, imparting radiation pressure. This allows the LG mode (a) to be changed without risking misalignment of the beam incident on the membrane. The integrating sphere lies on the other side of the membrane, creating a sensitive speckle pattern (b) which is recorded via a scientific camera. The light input into the integrating sphere is doubled to 532nm to improve speckle sensitivity and facilitate pattern acquisition. The acousto-optic modulator (AOM) allows the speckle pattern to be calibrated against wavelength deviations.}
    \label{fig:proposed_exp}
\end{figure*}

One possible means of measuring this effect is using recent advancements in speckle metrology. Speckle is a complex interference pattern that has demonstrated promise in metrological endeavours \cite{metzger_harnessing_2017,vijayakumar_implementation_2019,metzger_integrating_2016,facchin_measurement_2022,facchin_measuring_2023,facchin_wavelength_2021,bruce_overcoming_2019,cao_perspective_2017,raskatla_speckle-based_2022}. As integrating spheres have shown the ability to detect extremely small changes in path length, one could conceivably leverage this to create a highly sensitive force transducer. Integrating spheres are particularly well suited for this class of precision measurement experiments, as discussed by \textcite{facchin2024determining}. By separating a hemisphere and a mirror-like reflective membrane, one can create a virtual sphere from the reflection. This would act very similarly to the original technique, except allow for changing the effective size of the sphere, and consequently the path length, by deforming the membrane. Figure \ref{fig:proposed_exp} shows the layout we envision.

The proposal uses a spatial light modulator to shape the laser wavefront and a second laser to generate the speckle. According to the theory presented here, the changes in force incident on this membrane are on the order of $20$ fN per mode. The deflection of the membrane for a sheet of silver leaf ($500$ nm thickness) can be modelled, yielding a central deflection of approximately $0.28$ nm/fN. For a $50$ mm diameter hemisphere, this corresponds to an estimated deflection of $112$ ppb length change. \textcite{facchin_measuring_2023} measured $27$ pm displacements in a similar configuration using a $12.5$ mm radius sphere. This corresponds to 2.2 ppb of resolved length change, or over $50$ times smaller. This measurement therefore appears to be well within resolvable limits.

It is worth noting that such a precise measurement poses significant alignment and stability challenges. As described earlier, slight misalignment of the force-imparting beam—on the order of arcseconds—can overwhelm the measurement. Combined with other sources of environmental noise ubiquitous in precision measurements of a similar magnitude, such a measurement may prove difficult, but not impossible.

Another possible approach that has been suggested is to use a levitated particle within an optical trap to measure the radiation pressure of structured light. While levitated optomechanics has indeed proven effective in detecting small photon recoil forces \cite{jain2016direct}, this approach may not be as well suited for this measurement given the significant variation in beam size with mode index. Structured light beams, with complex spatial or polarisation distributions, are prone to inducing spin-orbit coupling effects on the particle. The extremely tight focusing of beams in optical tweezers experiments also presents strong field gradients and a potentially non-paraxial regime at the point of interest. The more gentle focusing in the experiment we propose is still able to resolve the effect, whilst avoiding any subtle measurement issues that such a complex environment may present.

This interaction between the light’s spin angular momentum (linked to polarisation) and orbital angular momentum (linked to spatial structure) introduces transverse or rotational forces that obscure a straightforward measurement of radiation pressure. Moreover, structured light fields can destabilise the particle within the trap, and optical resonances or scattering effects within the particle can produce additional radiative forces. These complicating factors would introduce spurious signals and make achieving the femtoNewton level force sensitivity required here challenging. In contrast, the hemispherical integrating sphere approach with speckle metrology offers a more stable and direct transduction method, circumventing these issues to provide a clearer pathway to precise radiation pressure measurements.

\section{Conclusions}
The nature of light has been a subject of interest since at least the 16th century. Understanding and leveraging both radiation pressure and the structure of light beams has facilitated numerous scientific advances in recent years. In this paper, we detail the subtle variations in radiation pressure exerted by structured light beams compared to plane waves, illustrating a strong dependence on the beam mode structure. This provides a direct link between the reduction of radiation pressure with higher-order modes and the Gouy phase of such beams. These findings not only advance our theoretical understanding of optomechanical interactions at microscales but also enhance our capability to manipulate mesoscopic particles with high precision. Despite the inherent limitations related to the required knowledge of beam mode structures, the experimental and theoretical approaches developed here lay a robust groundwork for further exploration into the nuanced dynamics of structured light. The connection to the beam propagation factor provides a means of applying this radiation pressure model to real beams with unknown mode structures. Expanding on this, we propose an experimental realisation that would allow for a potential means of measuring the radiation pressure effect in future research. This model provides insight into the subtle, but very real, connection between the structure of light and the corresponding radiation pressure it can exert on a surface.

\ack
We thank the Australian Research Council for funding (grant FL210100099). This research was supported by an Australian Government Research Training Program (RTP) Scholarship.

\printbibliography

@article{feng_physical_2001,
	title = {Physical origin of the Gouy phase shift},
	volume = {26},
	pages = {485},
	number = {8},
	journaltitle = {Optics Letters},
	shortjournal = {Opt. Lett.},
	author = {Feng, Simin and Winful, Herbert G.},
	date = {2001},
	langid = {english},
	}

@article{lembessis_mechanical_2016,
	title = {Mechanical effects on atoms interacting with highly twisted Laguerre-Gaussian light},
	volume = {94},
	pages = {043854},
	number = {4},
	journaltitle = {Physical Review A},
	shortjournal = {Phys. Rev. A},
	author = {Lembessis, V. E. and Babiker, M.},
	date = {2016},
	langid = {english},
	}

@article{saghafi_beam_1998,
	title = {The beam propagation factor for higher order Gaussian beams},
	volume = {153},
	rights = {https://www.elsevier.com/tdm/userlicense/1.0/},
	pages = {207--210},
	number = {4},
	journaltitle = {Optics Communications},
	shortjournal = {Optics Communications},
	author = {Saghafi, S. and Sheppard, C.J.R.},
	date = {1998},
	langid = {english},
	}

@article{subbarao_topological_1995,
	title = {Topological phase in Gaussian beam optics},
	volume = {20},
	pages = {2162},
	number = {21},
	journaltitle = {Optics Letters},
	shortjournal = {Opt. Lett.},
	author = {Subbarao, D.},
	date = {1995},
	langid = {english},
	}

@article{vijayakumar_implementation_2019,
	title = {Implementation of a speckle-correlation-based optical lever with extended dynamic range},
	volume = {58},
	pages = {5982},
	number = {22},
	journaltitle = {Applied Optics},
	shortjournal = {Appl. Opt.},
	author = {Vijayakumar, A. and Jayavel, D. and Muthaiah, M. and Bhattacharya, Shanti and Rosen, Joseph},
	date = {2019},
	langid = {english},
	}

@article{facchin_wavelength_2021,
	title = {Wavelength sensitivity of the speckle patterns produced by an integrating sphere},
	volume = {3},
	pages = {035005},
	number = {3},
	journaltitle = {Journal of Physics: Photonics},
	shortjournal = {J. Phys. Photonics},
	author = {Facchin, Morgan and Dholakia, Kishan and Bruce, Graham D},
	date = {2021},
	langid = {english},
	}

@article{facchin_measuring_2023,
	title = {Measuring picometre-level displacements using speckle patterns produced by an integrating sphere},
	volume = {13},
	pages = {14607},
	number = {1},
	journaltitle = {Scientific Reports},
	shortjournal = {Sci Rep},
	author = {Facchin, Morgan and Bruce, Graham D. and Dholakia, Kishan},
	date = {2023},
	langid = {english},
	}

@article{facchin_measurement_2022,
	title = {Measurement of Variations in Gas Refractive Index with 10 $^{\textrm{–9}}$ Resolution Using Laser Speckle},
	volume = {9},
	pages = {830--836},
	number = {3},
	journaltitle = {{ACS} Photonics},
	shortjournal = {{ACS} Photonics},
	author = {Facchin, Morgan and Bruce, Graham D. and Dholakia, Kishan},
	date = {2022},
	langid = {english},
	}

@article{facchin2024determining,
  title={Determining intrinsic sensitivity and the role of multiple scattering in speckle metrology},
  author={Facchin, Morgan and Khan, Saba N and Dholakia, Kishan and Bruce, Graham D},
  journal={Nature Reviews Physics},
  volume={6},
  number={8},
  pages={500--508},
  year={2024},
  publisher={Nature Publishing Group UK London}
}

@article{metzger_harnessing_2017,
	title = {Harnessing speckle for a sub-femtometre resolved broadband wavemeter and laser stabilization},
	volume = {8},
	pages = {15610},
	number = {1},
	journaltitle = {Nature Communications},
	shortjournal = {Nat Commun},
	author = {Metzger, Nikolaus Klaus and Spesyvtsev, Roman and Bruce, Graham D. and Miller, Bill and Maker, Gareth T. and Malcolm, Graeme and Mazilu, Michael and Dholakia, Kishan},
	date = {2017},
	langid = {english},
	}

@article{raskatla_speckle-based_2022,
	title = {Speckle-based deep learning approach for classification of orbital angular momentum modes},
	volume = {39},
	pages = {759},
	number = {4},
	journaltitle = {Journal of the Optical Society of America A},
	shortjournal = {J. Opt. Soc. Am. A},
	author = {Raskatla, Venugopal and Singh, B. P. and Patil, Satyajeet and Kumar, Vijay and Singh, R. P.},
	date = {2022},
	langid = {english},
	}

@inproceedings{metzger_integrating_2016,
	location = {Rochester, New York},
	title = {Integrating sphere based speckle generation for wavelength determination and laser stabilization},
	eventtitle = {Frontiers in Optics},
	pages = {FTh4C.4},
	booktitle = {Frontiers in Optics 2016},
	publisher = {{OSA}},
	author = {Metzger, Nikolaus Klaus and Spesyvtsev, Roman and Mazilu, Michael and Miller, Bill and Maker, Gareth T. and Malcolm, Graeme and Dholakia, Kishan},
	date = {2016},
	langid = {english},
	}

@article{hyde_m2_2019,
	title = {M2 factor of a vector Schell-model beam},
	volume = {58},
	pages = {1},
	number = {7},
	journaltitle = {Optical Engineering},
	shortjournal = {Opt. Eng.},
	author = {Hyde, Milo W. and Spencer, Mark F.},
	date = {2019},
	langid = {english},
	}

@inproceedings{siegman_how_1998,
	location = {Washington, D.C.},
	title = {How to (Maybe) Measure Laser Beam Quality},
	eventtitle = {Diode Pumped Solid State Lasers: Applications and Issues},
	pages = {MQ1},
	booktitle = {{DPSS} (Diode Pumped Solid State) Lasers: Applications and Issues},
	publisher = {{OSA}},
	author = {Siegman, A. E.},
	date = {1998},
	langid = {english},
	}

@article{phillips_spot_1983,
	title = {Spot size and divergence for Laguerre Gaussian beams of any order},
	volume = {22},
	pages = {643},
	number = {5},
	journaltitle = {Applied Optics},
	shortjournal = {Appl. Opt.},
	author = {Phillips, Ronald L. and Andrews, Larry C.},
	date = {1983},
	langid = {english},
	}

@article{stoyanov_gouy_2023,
	title = {Gouy phase of Bessel-Gaussian beams: theory vs. experiment},
	volume = {31},
	shorttitle = {Gouy phase of Bessel-Gaussian beams},
	pages = {13683},
	number = {9},
	journaltitle = {Optics Express},
	shortjournal = {Opt. Express},
	author = {Stoyanov, Lyubomir and Stefanov, Aleksander and Dreischuh, Alexander and Paulus, Gerhard G.},
	date = {2023},
	langid = {english},
	}

@article{davis_theory_1979,
	title = {Theory of electromagnetic beams},
	volume = {19},
	pages = {1177--1179},
	number = {3},
	journaltitle = {Physical Review A},
	shortjournal = {Phys. Rev. A},
	author = {Davis, L. W.},
	date = {1979},
	langid = {english},
	}

@article{rohrbach_switching_2005,
	title = {Switching and measuring a force of 25 {femtoNewtons} with an optical trap},
	volume = {13},
	pages = {9695},
	number = {24},
	journaltitle = {Optics Express},
	shortjournal = {Opt. Express},
	author = {Rohrbach, Alexander},
	date = {2005},
	langid = {english},
	}

@article{wan_propagation_2023,
	title = {The propagation speed of optical speckle},
	volume = {13},
	pages = {9071},
	number = {1},
	journaltitle = {Scientific Reports},
	shortjournal = {Sci Rep},
	author = {Wan, Zhenyu and Yessenov, Murat and Padgett, Miles J.},
	date = {2023},
	langid = {english},
	}

@inproceedings{li_traceable_2019,
	location = {Berlin, Germany},
	title = {Traceable Laser Power Measurement Using a Micro-Machined Force Sensor with Sub-Piconewton Resolution},
	eventtitle = {2019 20th International Conference on Solid-State Sensors, Actuators and Microsystems \& Eurosensors {XXXIII} ({TRANSDUCERS} \& {EUROSENSORS} {XXXIII})},
	pages = {1603--1606},
	booktitle = {2019 20th International Conference on Solid-State Sensors, Actuators and Microsystems \& Eurosensors {XXXIII} ({TRANSDUCERS} \& {EUROSENSORS} {XXXIII})},
	publisher = {{IEEE}},
	author = {Li, Zhi and Gao, Sai and Brand, Uwe and Hiller, Karla and Hahn, Susann and Wolff, Helmut},
	date = {2019},
	langid = {english},
	}

@article{shaw_comparison_2019,
	title = {Comparison of electrostatic and photon pressure force references at the nanonewton level},
	volume = {56},
	pages = {025002},
	number = {2},
	journaltitle = {Metrologia},
	shortjournal = {Metrologia},
	author = {Shaw, Gordon A and Stirling, Julian and Kramar, John and Williams, Paul and Spidell, Matthew and Mirin, Richard},
	date = {2019},
	langid = {english},
	}

@article{qiu_directly_2021,
	title = {Directly observing the skew angle of a Poynting vector in an {OAM} carrying beam via angular diffraction},
	volume = {46},
	pages = {3484},
	number = {14},
	journaltitle = {Optics Letters},
	shortjournal = {Opt. Lett.},
	author = {Qiu, Song and Ren, Yuan and Liu, Tong and Liu, Zhengliang and Wang, Chen and Ding, You and Sha, Qimeng and Wu, Hao},
	date = {2021},
	langid = {english},
	}

@article{simpson_mechanical_1997,
	title = {Mechanical equivalence of spin and orbital angular momentum of light: an optical spanner},
	volume = {22},
	number = {1},
	journaltitle = {Optics Letters},
	author = {Simpson, N B and Dholakia, K and Allen, L and Padgett, M J},
	date = {1997},
	langid = {english},
	}

@article{padgett_orbital_2017,
	title = {Orbital angular momentum 25 years on [Invited]},
	volume = {25},
	pages = {11265},
	number = {10},
	journaltitle = {Optics Express},
	shortjournal = {Opt. Express},
	author = {Padgett, Miles J.},
	date = {2017},
	langid = {english},
	}

@article{leach_direct_2006,
	title = {Direct measurement of the skew angle of the Poynting vector in a helically phased beam},
	volume = {14},
	pages = {11919},
	number = {25},
	journaltitle = {Optics Express},
	shortjournal = {Opt. Express},
	author = {Leach, Jonathan and Keen, Stephen and Padgett, Miles J. and Saunter, Christopher and Love, Gordon D.},
	date = {2006},
	langid = {english},
	}

@article{chen_dynamics_2013,
	title = {Dynamics of microparticles trapped in a perfect vortex beam},
	volume = {38},
	pages = {4919},
	number = {22},
	journaltitle = {Optics Letters},
	shortjournal = {Opt. Lett.},
	author = {Chen, Mingzhou and Mazilu, Michael and Arita, Yoshihiko and Wright, Ewan M. and Dholakia, Kishan},
	date = {2013},
	langid = {english},
	}

@article{cao_perspective_2017,
	title = {Perspective on speckle spectrometers},
	volume = {19},
	pages = {060402},
	number = {6},
	journaltitle = {Journal of Optics},
	shortjournal = {J. Opt.},
	author = {Cao, Hui},
	date = {2017},
	langid = {english},
	}

@article{bruce_overcoming_2019,
	title = {Overcoming the speckle correlation limit to achieve a fiber wavemeter with attometer resolution},
	volume = {44},
	pages = {1367},
	number = {6},
	journaltitle = {Optics Letters},
	shortjournal = {Opt. Lett.},
	author = {Bruce, Graham D. and O’Donnell, Laura and Chen, Mingzhou and Dholakia, Kishan},
	date = {2019},
	langid = {english},
	}

@article{beth_mechanical_1936,
	title = {Mechanical Detection and Measurement of the Angular Momentum of Light},
	volume = {50},
	rights = {http://link.aps.org/licenses/aps-default-license},
	pages = {115--125},
	number = {2},
	journaltitle = {Physical Review},
	shortjournal = {Phys. Rev.},
	author = {Beth, Richard A.},
	date = {1936},
	langid = {english},
	}

@article{lebedew_untersuchungen_1901,
	title = {Untersuchungen über die Druckkräfte des Lichtes},
	volume = {311},
	pages = {433--458},
	number = {11},
	journaltitle = {Annalen der Physik},
	shortjournal = {Annalen der Physik},
	author = {Lebedew, Peter},
	date = {1901},
	langid = {german},
	}

@article{nichols_pressure_1903,
	title = {The Pressure due to Radiation},
	volume = {17},
	pages = {315--351},
	number = {5},
	journaltitle = {The Astrophysical Journal},
	author = {Nichols, E. F. and Hull, G. F.},
	date = {1903},
	}

@article{nichols_preliminary_1901,
	title = {A Preliminary Communication on the Pressure of Heat and Light Radiation},
	volume = {13},
	pages = {307--320},
	number = {5},
	journaltitle = {Physical Review (Series I)},
	shortjournal = {Phys. Rev. (Series I)},
	author = {Nichols, E. F. and Hull, G. F.},
	date = {1901},
	langid = {english},
	}

@article{poynting_wave_1909,
	title = {The wave motion of a revolving shaft, and a suggestion as to the angular momentum in a beam of circularly polarised light},
	author = {Poynting, J. H.},
	date = {1909},
	langid = {english},
	}

@article{einstein_uber_1905,
	title = {Über einen die Erzeugung und Verwandlung des Lichtes betreffenden heuristischen Gesichtspunkt},
	volume = {322},
	pages = {132--148},
	number = {6},
	journaltitle = {Annalen der Physik},
	shortjournal = {Annalen der Physik},
	author = {Einstein, A.},
	date = {1905},
	langid = {german},
	}

@article{christianson_tycho_1979,
	title = {Tycho Brahe's German Treatise on the Comet of 1577: A Study in Science and Politics},
	volume = {70},
	shorttitle = {Tycho Brahe's German Treatise on the Comet of 1577},
	pages = {110--140},
	number = {1},
	journaltitle = {Isis},
	shortjournal = {Isis},
	author = {Christianson, J. R. and Brahe, Tycho},
	date = {1979},
	langid = {english},
	}

@article{carter_spot_1980,
	title = {Spot size and divergence for Hermite Gaussian beams of any order},
	volume = {19},
	pages = {1027--1029},
	number = {7},
	journaltitle = {Optica Publishing Group},
	author = {Carter, William H},
	date = {1980},
	}

@article{jain2016direct,
  title={Direct measurement of photon recoil from a levitated nanoparticle},
  author={Jain, Vijay and Gieseler, Jan and Moritz, Clemens and Dellago, Christoph and Quidant, Romain and Novotny, Lukas},
  journal={Physical review letters},
  volume={116},
  number={24},
  pages={243601},
  year={2016},
  publisher={APS}
}

\end{document}